\documentclass{appolb}
\usepackage{graphicx}

\begin{document}
\title{Recent results from NA61/SHINE%
\thanks{Presented at Matter To The Deepest
Recent Developments In Physics Of Fundamental Interactions
XXXIX International Conference of Theoretical Physics }%
}
\author{Szymon Pulawski\\ for the NA61/SHINE collaboration
\address{Institute of Physics, University of Silesia, Katowice, Poland}
}
\maketitle
\begin{abstract}
The main physics goals of the NA61/SHINE programme on strong interactions are the study of the properties of the onset of deconfinement and the search for signatures of the critical point of strongly interacting matter. These goals are pursued by performing an energy (beam momentum 13$A$ -- 158$A$~GeV/c) and system size (p+p, p+Pb, Be+Be, Ar+Sc, Xe+La) scan.\\
This publication reviews results and plans of NA61/SHINE. In particular, recent inclusive spectra and new results on fluctuations and correlations of identified hadrons in inelastic p+p and centrality selected Be+Be interactions at the SPS energies are presented. The energy dependence of quantities inspired by the Statistical Model of the Early Stage (kink, horn and step) show unexpected behaviour in p+p collisions. The NA61/SHINE results are compared with the corresponding data of other experiments and model predictions.\end{abstract}
\PACS{PACS numbers come here}
  
\section{The NA61/SHINE facility}
\begin{figure}[htb]
\centerline{%
\includegraphics[width=12.5cm]{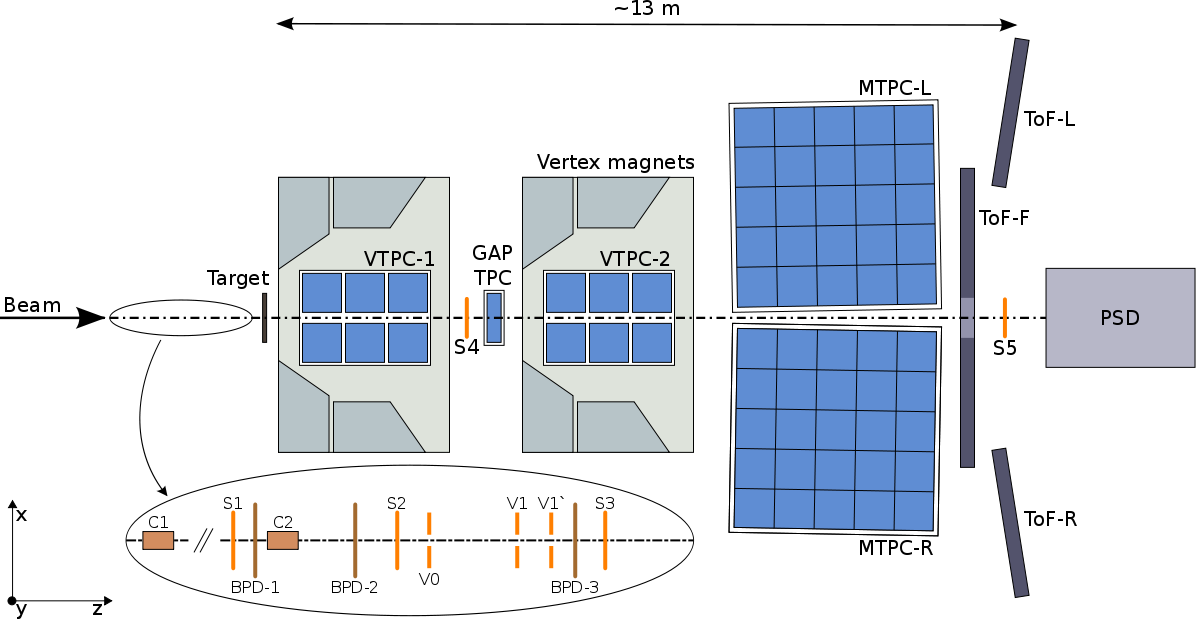}}
\caption{Schematic view of the NA61/SHINE detector system.}
\label{fig:detector}
\end{figure}

The layout of the NA61/SHINE detector~\cite{detectorpaper} is presented in Figure~\ref{fig:detector}. It consists of a large
acceptance hadron spectrometer with excellent capabilities in charged particle momentum measurements
and identification by a set of five Time Projection Chambers as well as Time-of-Flight
detectors. The high resolution forward calorimeter, the Projectile Spectator Detector, measures energy
flow around the beam direction, which in nucleus-nucleus reactions is primarily a measure of
the number of spectator (non-interacted) nucleons and is thus related to the centrality of the collision.
Additionally, detector system consist of three Time-of-Flight walls used for low momentum particle identification. An array of beam detectors identifies beam particles, secondary hadrons and ions as well as
primary ions, and measures precisely their trajectories. 

The NA61/SHINE experiment up to now used secondary hadron beams in the momentum range from 13 GeV/c to 350 GeV/c, as well as attenuated primary Pb$^{82+}$ ion beams in the momentum range from 13$A$ GeV/c to 158$A$ GeV/c. For the start of the strong interactions physics program secondary $^{7}$Be ions were produced via fragmentation of the Pb$^{82+}$ ions as primary ions other than Pb were not available before 2015. 
In 2015 primary $^{40}$Ar ion beams in the same momentum range were delivered to NA61/SHINE and $^{131}$Xe beams are expected in 2017.

\section{Results from $^7$Be+$^9$Be and p+p collisions}
\subsection{Inelastic $^7$Be+$^9$Be cross section}
\begin{figure}[htb]
\centerline{%
\includegraphics[width=8.5cm]{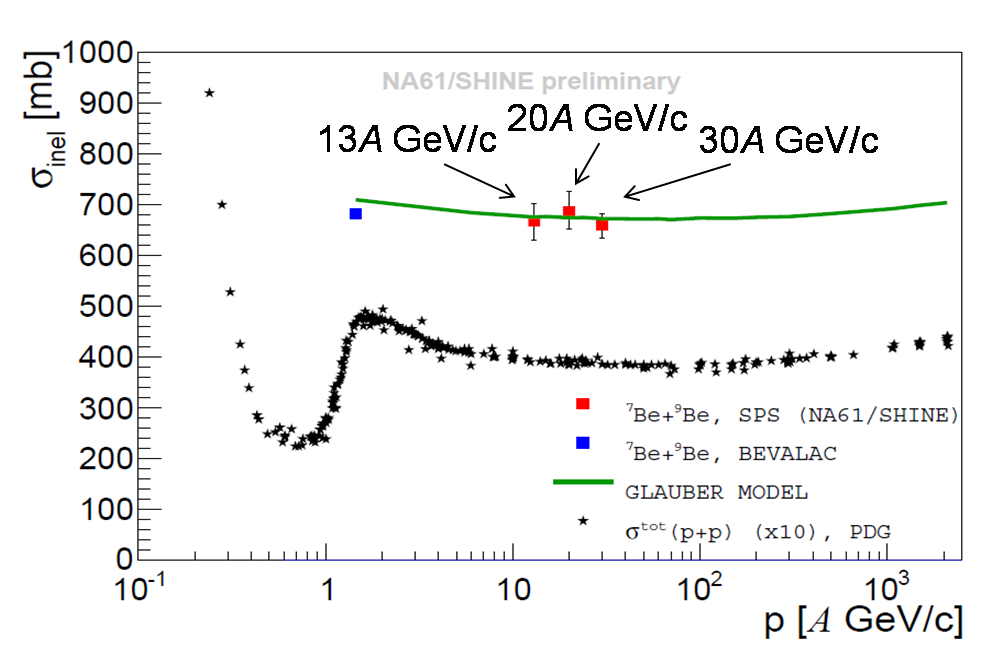}}
\caption{Energy dependence of the inelastic $^7$Be+$^9$Be cross section as a function of beam momentum. Results are compared with the measurement of Ref.~\cite{sigmalow}, world p+p data and Glissando model predictions.}
\label{fig:inelastic}
\end{figure}

In the data taking on $^7$Be+$^9$Be collisions a 2 cm diameter interaction trigger
counter S4 was placed on the beam-line between VTPC-1 and VTPC-2. This arrangement allowed to measure the total
inelastic $^7$Be+$^9$Be cross section $\sigma_{inel}$~\cite{sigmain}. The NA61/SHINE measurements at 13$A$, 20$A$ and
30$A$ GeV/c are presented in Figure~\ref{fig:inelastic}. The NA61/SHINE results are in good agreement with an earlier
measurement at lower beam momentum~\cite{sigmalow} and a Glauber model calculation using the Glissando code~\cite{Glisando}.

\subsection{Rapidity distributions}
\begin{figure}[htb]
\centerline{%
\includegraphics[width=10.5cm]{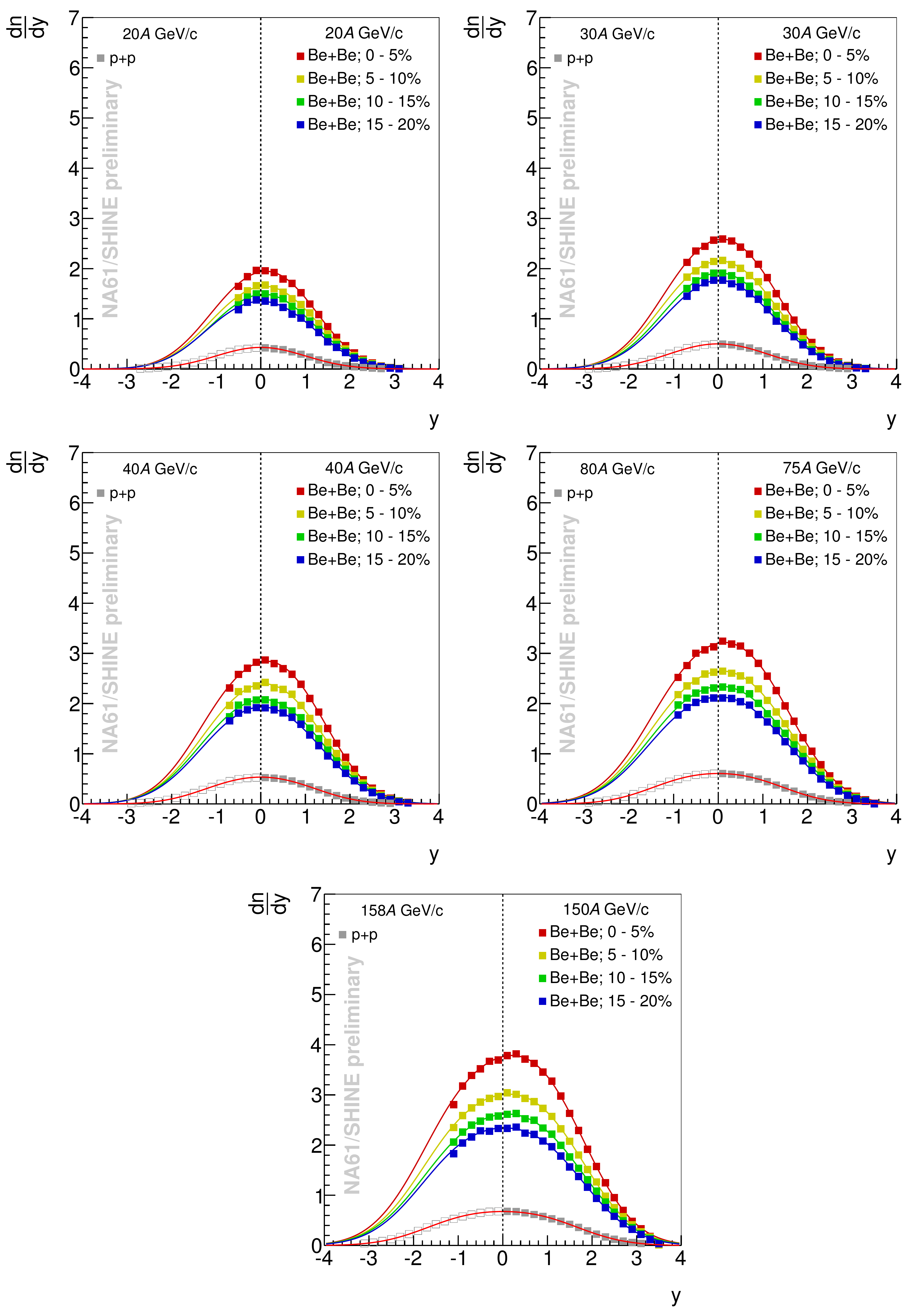}}
\caption{$\pi^{-}$ rapidity distributions of $\pi^-$ produced in inelastic p+p and Be+Be interactions (4 centrality classes) at 20$A$, 30$A$, 40$A$, 75$A$ and 150$A$ GeV/c.}
\label{fig:rapidity}
\end{figure}

The rapidity spectra of $\pi^-$ in $^7$Be+$^9$Be collisions at the three beam momenta and four centrality
classes together with data for inelastic p+p interactions are presented in Figure~\ref{fig:rapidity}. One can observe
a small asymmetry in the rapidity distribution for $^7$Be+$^9$Be collisions around mid-rapidity. This
asymmetry may come from two effects:
\begin{itemize}
\item the asymmetry between projectile and target nuclei ($^7$Be projectile on $^9$Be target) which is
expected to enhance particle production in the backward (target) hemisphere,
\item the selection of central collisions which requires a small number of projectile spectators
without any explicit requirement imposed on the number of target spectators; this selection, when used for collisions of identical nuclei, would enhance particle production in the forward hemisphere.
\end{itemize}
Note that the two effects partially compensate leading to a relatively small asymmetry of
the measured spectra.

\subsection{Particle ratios and inverse slope parameter~T of $m_T$ distributions in inelastic p+p collisions}
\begin{figure}[htb]
\centerline{%
\includegraphics[width=12.5cm]{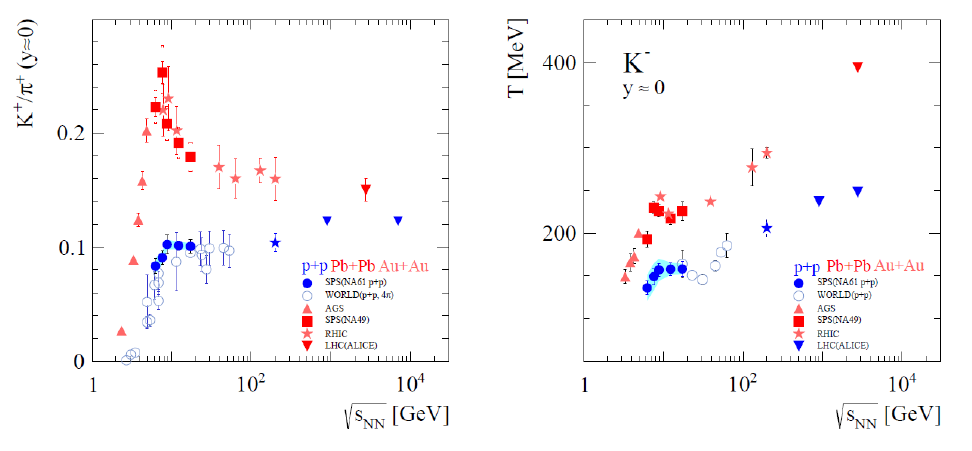}}
\caption{$K^+/\pi^+$ ratio and inverse slope parameter T of $K^{-}$ transverse mass spectra at mid-rapidity.}
\label{fig:horn}
\end{figure}

The excellent particle identification based on the combined time-of-flight (tof) and energy loss (dE/dx) method allow us to calculate
$K^{+}/\pi^+$ ratios. The energy dependence of the $K^+/\pi^+$ ratio at mid-rapidity
for inelastic p+p interactions and central Pb+Pb/Au+Au collisions is presented in Figure~\ref{fig:horn}.
The NA61/SHINE data suggest that even in inelastic p+p interactions the energy dependence of the $K^+/\pi^+$ ratio exhibits rapid changes in the SPS energy range. A step structure is seen which appears to be a precursor of the horn structure~\cite{6} observed in central Pb+Pb collisions. Data obtained at RHIC and LHC~\cite{7,8,9,10,11} are also plotted in Figure~\ref{fig:horn} to show the trend beyond the SPS energy range.

The inverse slope parameter T was fitted to the transverse mass spectra. Figure~\ref{fig:horn} presents the energy dependence of T for $K^-$ in inelastic p+p reactions from  NA61/SHINE compared to world data for p+p and Pb+Pb/Au+Au reactions from Refs. \cite{10,12,13,14}.
Results from inelastic p+p collisions exhibit a step structure like the one observed in central Pb+Pb interactions.

\subsection{$\Lambda$ hyperon production in inelastic p+p interactions at 158 GeV/c}
\begin{figure}[htb]
\centerline{%
\includegraphics[width=12.5cm]{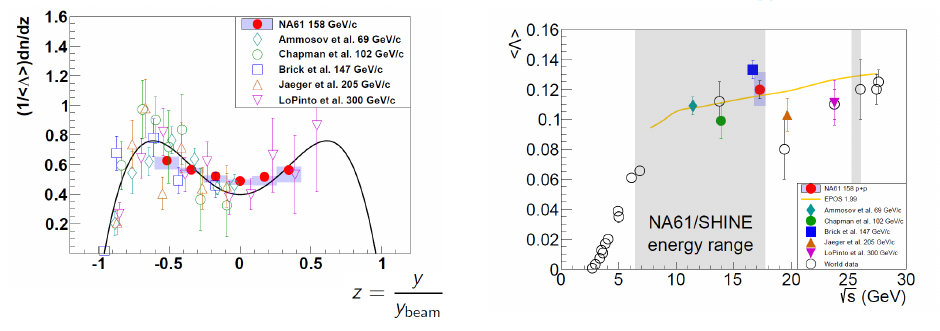}}
\caption{Rapidity distribution of $\Lambda$ produced in inelastic p+p collisions at 158 GeV/c.}
\label{fig:Lambda}
\end{figure}

The $\Lambda$ rapidity distribution measured by NA61/SHINE is compared to previous measurements
in Figure~\ref{fig:Lambda}. In order to partly compensate for differences resulting from different collision energies
the comparison is done in terms of the scaled rapidity $z = y/y_{beam}$ and the spectra are normalized
to unity. The mean $\Lambda$ multiplicity is calculated by summing up the measured rapidity
spectrum and extrapolating it to lower and higher rapidities using the parametrization of the world
data from Figure~\ref{fig:Lambda}. The collision energy dependence of the mean $\Lambda$ multiplicity including the
new NA61/SHINE result is plotted in Figure~\ref{fig:Lambda}.

\subsection{Transverse momentum fluctuations in $^7$Be+$^9$Be collisions}
The  transverse momentum fluctuations were studied using strongly intensive measure $\Sigma[P_{T},N]$~\cite{Tobiasz} for $^7$Be+$^9$Be interactions at several collision centralities and energies. The results were compared to those obtained from inelastic p+p interactions by NA61/SHINE and are very similar.  Transverse momentum fluctuations in $^7$Be+$^9$Be interactions do not present structures which could be related to the Critical Point.

\section{Acknowledgments}
This work was supported by the National Science Center of Poland (grants: 2014/12/T/ST2/00692, 2013/11/N/ST2/03879, 2012/04/M/ST2/00816). 
%


\end{document}